\documentclass[amssymb,amsmath,prd,twocolumn,superscriptaddress,nofootinbib]{revtex4-1}
\usepackage[utf8]{inputenc}

\usepackage{graphicx}
\usepackage{bm}
\usepackage{amssymb,amsmath}
\usepackage{mathrsfs}
\usepackage{latexsym}
\usepackage{color}
\usepackage[normalem]{ulem}
\usepackage{dcolumn}
\usepackage[colorlinks=true,citecolor=blue,urlcolor=blue]{hyperref}
\usepackage[usenames,dvipsnames]{xcolor}
\usepackage{xspace}
\usepackage{graphicx}
\usepackage{float}
\usepackage{pifont}% http://ctan.org/pkg/pifont
\newcommand{\cmark}{\ding{51}}%
\newcommand{\xmark}{\ding{55}}%
\usepackage{multirow}

\newcommand{\chieff}{\ensuremath{\chi_{\rm eff}}\xspace}
\newcommand{\Mc}{\ensuremath{{\cal{M}}\xspace}}
\newcommand{\DL}{\ensuremath{D_{\rm L}\xspace}}
\definecolor{shgreen}{rgb}{0.15625, 0.609375, 0.316406}

\newcommand{\CIT}{\affiliation{Department of Physics, California Institute of Technology, Pasadena, California 91125, USA}}
\newcommand{\CITLab}{\affiliation{LIGO Laboratory, California Institute of Technology, Pasadena, California 91125, USA}}
\newcommand{\Amherst}{\affiliation{Department of Physics and Astronomy, Amherst College, Amherst, Massachusetts 01002, USA}}

\begin{document}
\title{Concurrent estimation of noise and compact-binary signal parameters in gravitational-wave data}
\author{Cailin Plunkett} \thanks{cplunkett23@amherst.edu} \Amherst
\author{Sophie Hourihane} \CIT
\author{Katerina Chatziioannou} \CIT \CITLab 

\date{\today}

\begin{abstract}
    Gravitational-wave parameter estimation for compact binary signals typically relies on sequential estimation of the properties of the detector Gaussian noise and of the binary parameters. This procedure assumes that the noise variance, expressed through its power spectral density, is perfectly known in advance. We assess the impact of this approximation on the estimated parameters by means of an  analysis that simultaneously estimates the noise and compact binary parameters, thus allowing us to marginalize over uncertainty in the noise properties. We compare the traditional sequential estimation method and the new full marginalization method using events from the GWTC-3 catalog. We find that the recovered signals and inferred parameters agree to within their statistical measurement uncertainty. At current detector sensitivities, uncertainty about the noise power spectral density is a subdominant effect compared to other sources of uncertainty.
\end{abstract}

\maketitle

%%%%%%%%%%%%%%%%%%%%%%%%%%%%%%%%%%%%%%%%%%
\section{Introduction}
%%%%%%%%%%%%%%%%%%%%%%%%%%%%%%%%%%%%%%%%%%

Estimating the source properties of compact binary coalescences (CBC) using data from gravitational-wave (GW) observatories~\cite{LIGOScientific:2014pky, VIRGO:2014yos} relies on accurate models of both the astrophysical signal and the noise in the detector. While the signal model employs waveform templates that are based on analytic approximations to Einstein's field equations, the noise model typically takes the form of the power spectral density (PSD) of the detector Gaussian noise, $S_n(f)$~\cite{LIGOScientific:2019hgc}. 
Given GW data $d=h+n$ that are a sum of a coherent signal $h$ between detectors and uncorrelated random noise $n$, and an accurate signal model $h'$, the residual $r=d-h'$ should follow the same distribution as the noise. Assuming Gaussian and stationary noise, the likelihood function $\mathcal L(d|h')$ in the frequency domain is~\cite{Romano:2016dpx,Cornish2021}
\begin{equation}
    \mathcal L(d|h') = \prod_i \frac{2}{\pi T S_n(f_i)} \exp{\left[-2\frac{|\tilde{r_i}|^2}{TS_n(f_i)}\right]}\,,
    \label{eq:likelihood}
\end{equation}
which explicitly depends on $S_n(f)$; $T$ is the duration of the data analyzed and $i$ counts frequency bins. 

The likelihood of Eq.~\eqref{eq:likelihood}, and thus analyses of GW signals, depend on multiple formal assumptions about the detector noise and its PSD. Parameter estimation~\cite{Veitch2015,Ashton2019,Romero-Shaw:2020owr} assumes that the noise is Gaussian and stationary in time, though mitigation techniques and methods to relax these assumptions have been explored~\cite{Zackay:2019kkv, Chatziioannou:2021ezd,Hourihane:2022doe,Cornish:2020odn,2021PhRvD.103l4061E}. Furthermore, the PSD is typically a required input for CBC parameter estimation. Using a point estimate for $S_n(f)$ assumes the noise PSD is perfectly known or at least that is it estimated to precision far greater than other statistical uncertainties. 
The noise PSD must be estimated from the data~\cite{LIGOScientific:2019hgc}, typically in one of two ways. The first, termed the ``off-source'' method, relies on the median PSD of many data segments around, but not including, the data segment containing the signal~\cite{Veitch2015}. This calculation requires the stationarity assumption to hold over the entire stretch of data used to estimate the PSD, typically ${\cal{O}}(10^3\mathrm{s})$~\cite{Veitch2015,LIGOScientific:2013yzb}. The second, termed the ``on-source'' method, uses only the data segment containing the signal.

The ``on-source" method relies on {\tt BayesLine}~\cite{Littenberg_2015}, integrated into the broader {\tt BayesWave} algorithm~\cite{Cornish_2015,Cornish:2020odn}, and models the PSD with a two-component phenomenological fit. A cubic spline describes the broadband noise behavior, while narrow-band spectral features are fit using Lorentzians. The number and parameters of the spline points and Lorentzians are marginalized over with a trans-dimensional Markov chain Monte Carlo (MCMC) algorithm~\cite{10.1093/biomet/82.4.711}, also known as a reversible-jump MCMC (RJMCMC). 
Any non-Gaussian components of the data (namely, instrumental glitches or astrophysical signals) are modeled with sine-Gaussian wavelets, thus relaxing the assumption that the noise is Gaussian during PSD estimation. By using the same data segment for both PSD and signal parameter estimation, the on-source method additionally restricts the stationarity assumption to a shorter data segment, typically ${\cal{O}}(4\mathrm{s})$. A point estimate from the PSD posterior, typically a fair draw or the median, is then selected and passed on for downstream analyses.
The on-source and off-source methods were compared in~\cite{Chatziioannou_2019} where it was demonstrated that the former results in PSDs that are more consistent with the Gaussian likelihood assumptions. Both methods, however, still provide a single point-estimate for the PSD.

Traditional parameter estimation~\cite{Abbott2019events,AbbottO3aEvents,LIGOScientific:2021djp} involves sequential estimation of the noise PSD and the binary parameters. Once a point-estimate for the noise PSD has been provided with either the on-source or the off-source methods, {\tt LALInference}~\cite{Veitch2015}, {\tt Bilby}~\cite{Ashton2019,Romero-Shaw:2020owr}, or other similar pipelines marginalize over the binary parameters using waveform templates as the signal model. During this two-step sequential process, care must be taken to ensure that the spectral resolution and maximum and minimum frequencies of the estimated PSD match the desired parameter estimation analysis settings. Besides these technical complications, the sequential process assumes that the PSD is perfectly estimated with negligible uncertainty, or at least that any uncertainty is not correlated with the binary parameters.

Assessing the impact of the assumption that the noise PSD is perfectly known requires marginalizing over noise uncertainty during parameter estimation. The choice of PSD estimation method and PSD prior results in different functional forms for the marginalized CBC likelihood~\cite{Rover2010,Rover:2011qd,2013PhRvD..88h4044L,Littenberg_2015,Edwards:2015eka,Banagiri:2019lon,Smith:2017vfk,Biscoveanu2020,Talbot2020}.~\citet{Rover2010} showed that assuming a PSD prior that is an inverse $\chi^2$-distribution allows one to analytically marginalize over PSD uncertainty and converts the standard Whittle likelihood to a Student-t distribution.  \citet{Talbot2020} extended~\citep{Rover2010} and computed the marginalized likelihood for both the mean and median off-source PSD. Finally,~\citet{Biscoveanu2020} explored a method of incorporating PSD uncertainty in CBC parameter estimation by considering the full PSD posterior from {\tt BayesLine} rather than a point-estimate. They still analyzed the data sequentially, but used $200$ fair draws from the PSD posterior to perform $200$ parameter estimation analyses and combined the resulting posteriors. Besides the computational cost, one drawback of this method is that it ignores potential correlations between the PSD and the CBC parameter posterior.

Both~\citet{Biscoveanu2020} and~\citet{Talbot2020} concluded that PSD uncertainty is likely a subdominant effect compared to typical current statistical uncertainties given their respective PSD and binary parameter estimation methods. \citet{Biscoveanu2020} found posteriors under the median and marginalized on-source PSD that agree in position and width to within a few percent, which they determine is an order of magnitude higher than expected due to statistical fluctuations. Among the GWTC-1 events~\cite{Abbott2019events}, GW151012~\cite{LIGOScientific:2016dsl} displays the largest effect due to uncertainty on the LIGO Hanford PSD. However, the choice of PSD estimation method, specifically the off-source mean or median studied in~\cite{Talbot2020}, had a greater impact on the inferred parameter posterior than PSD marginalization. Similar conclusions were reached in~\cite{Chatziioannou:2021tdi} in the context of the impact of PSD misestimation on inferring the tidal deformability of neutron star binaries. The impact of marginalization could be more significant when precise calculations of Bayesian evidence are needed~\cite{Talbot2020}.  

In this study we revisit the issue of uncertainty in the PSD estimation and its effect on the inferred CBC parameters on events from GWTC-3~\cite{LIGOScientific:2021djp}. We take advantage of recent developments on the {\tt BayesWave} algorithm\footnote{From now on we use the term {\tt BayesWave} to describe the combined {\tt BayesLine} and {\tt BayesWave} algorithms.} that augment its list of models with a CBC model that describes the signal with waveform templates~\cite{Chatziioannou:2021ezd,Wijngaarden:2022sah}. This allows us to \emph{simultaneously} model the noise PSD and the GW signal from the same on-source data, thus obtaining CBC parameter posteriors that are marginalized over noise uncertainty. Compared to the approach of~\cite{Biscoveanu2020}, our method allows us to obtain a joint posterior for the noise and the signal rather than assume the signal and noise posteriors are uncorrelated and obtained by sequential analysis of the same data.

We apply our analysis to binary black hole (BBH) signals from GWTC-3~\cite{LIGOScientific:2021djp}. To reduce computational cost, we restrict to confident detections with signal-to-noise ratio (SNR) $>10$. We further discard signals with analysis segments $>16$ seconds and events with technical complications of different lower frequency cutoffs in different detectors, for a total sample of $50$ events. We analyze each event with two methods. The \emph{sequential} method follows the procedure employed for parameter estimation in~\cite{LIGOScientific:2021djp} by first estimating the median for the on-source noise PSD and then using this point estimate to sample from the posterior of the CBC source parameters. The \emph{marginalized} method samples from the joint posterior of the noise PSD and CBC source parameters. We compare the resulting source parameter posteriors and find broad agreement between the two methods. In particular, over all events the marginalized method results in an average change of $0.0^{+1.4}_{-1.6}\%$ in the marginalized chirp mass posterior median and a decrease of $2^{+16}_{-13}\%$ in the marginalized posterior 90\% width. Neither are statistically distinguishable from zero. In addition, we find no significant trend between the change in posterior width and the posterior median or signal SNR. Finally, we devise a simple toy model to demonstrate how a broadband increase or reduction in the noise PSD results in a higher-order, and thus subdominant, effect in the CBC parameter posteriors.  

The rest of the paper is organized as follows. In Sec.~\ref{sec:methods}, we describe the {\tt BayesWave} algorithm and models for the GW signal available within it. In Sec.~\ref{sec:analyses}, we discuss our analyses on a set of events from GWTC-3~\cite{LIGOScientific:2021djp}, considering the waveform reconstructions as well as the parameter posteriors. We use a toy model of a shifted PSD to help explain the minimal impact of PSD marginalization on parameter posteriors in Sec.~\ref{sec:toymodel}. Finally, in Sec.~\ref{sec:conclusion} we conclude.  

%%%%%%%%%%%%%%%%%%%%%%%%%%%%%%%%%%%%%%%%%%
\section{Methods and Analysis} 
\label{sec:methods}
%%%%%%%%%%%%%%%%%%%%%%%%%%%%%%%%%%%%%%%%%%

In this section, we describe the features of {\tt BayesWave} that are relevant to our study as well as details about our analysis of GWTC-3, for which we use data from the Gravitational Wave Open Science Center (GWOSC)~\cite{Vallisneri:2014vxa,LIGOScientific:2019lzm}.

%----------------------------------------------
\subsection{Algorithm description}
\label{sec:alg}
%----------------------------------------------

%
\begin{figure}
    \centering
     \includegraphics[width=0.48\textwidth]{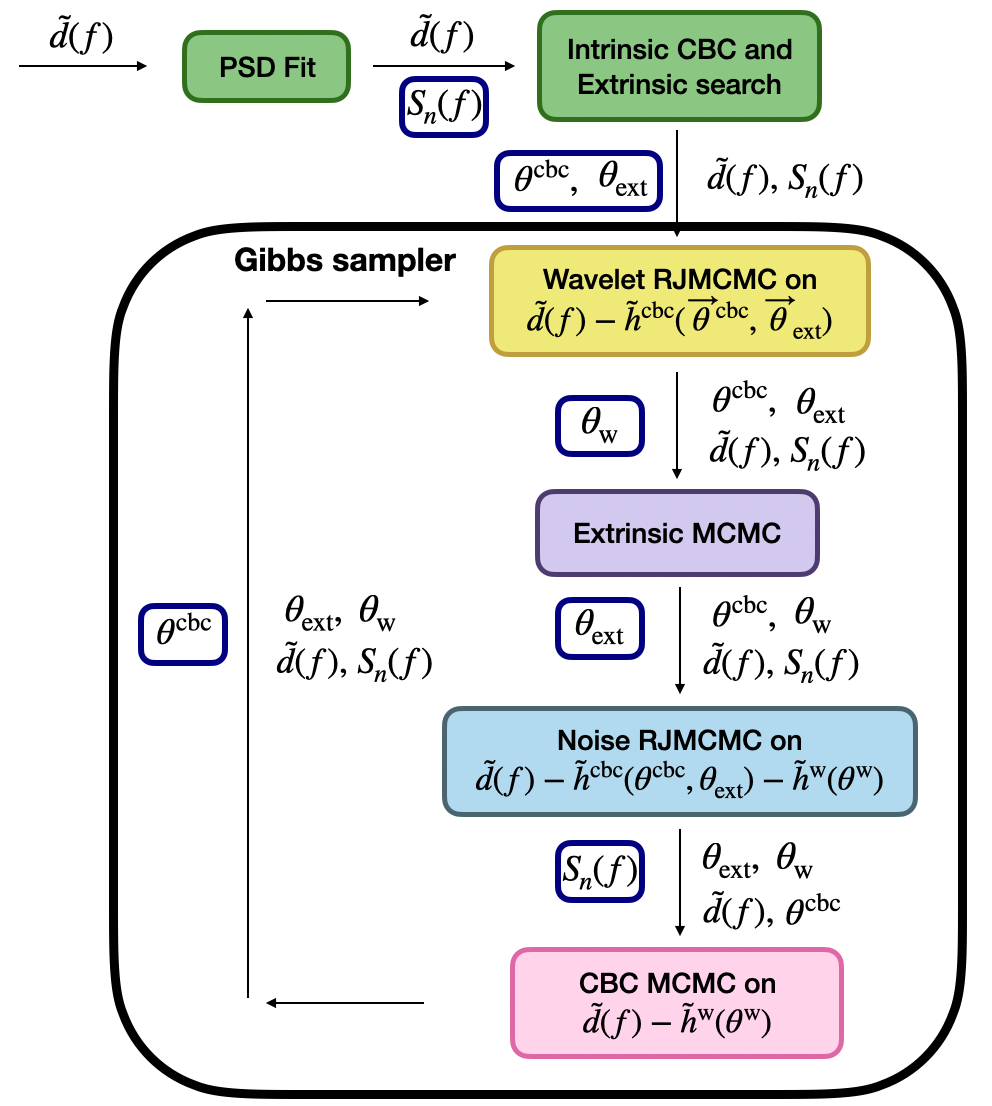}
    \caption{{\tt BayesWave} workflow diagram. Preconditioned frequency-domain data $\tilde{d}(f)$ undergo quick initial parameter fits to use as starting points for the blocked Gibbs sampler (black box). The Gibbs sampler cycles through the component samplers, which are given in filled colored boxes with their names and input parameters, using independent MCMCs. The yellow (purple, blue, pink) box represents the glitch (extrinsic parameter, noise PSD, intrinsic CBC parameter) sampler. Below each component sampler, the parameters that are held fixed (sampled) are unboxed (boxed in navy).}
    \label{fig:workflow}
\end{figure}
\begin{table*}[]
\begin{tabular}{l @{\quad\quad} c @{\quad\quad}c @{\quad}c @{\quad}c}
 \multirow{2}{*}{Model name} & \multirow{2}{*}{Model configuration} &  \multicolumn{3}{c}{Analysis mode}  \\
\cline{3-5}
& & Cleaning Phase & CBC+noise & CBC+fixed\\
\hline
\emph{CBC model}  & GW signal with CBC templates & \xmark & \cmark & \cmark\\ 
\emph{glitch model}  & Incoherent power with wavelets & \cmark & \xmark & \xmark\\ 
\emph{noise model}  & Noise PSD with splines and Lorentzians & \cmark &\cmark & \xmark\\ 
\emph{signal model}  & GW signal with wavelets  & \xmark & \xmark & \xmark\\ 
\hline
\hline
\end{tabular}
\caption{{\tt BayesWave} term glossary. The first column gives the model name, while the second column provides a brief description of its configuration. The third, fourth, and fifth columns correspond to the three analyses modes used in this study and denote which models are active during each.}
\label{tab:glossary}
\end{table*}

The {\tt BayesWave} algorithm is described in detail in~\cite{Cornish_2015,Cornish2021,Chatziioannou:2021ezd,Wijngaarden:2022sah} and here we describe the aspects relevant to this work. {\tt BayesWave} simultaneously models GW signals, detector noise, and glitches using different models. We provide a glossary of terms in Table~\ref{tab:glossary}, a schematic algorithm workflow in Fig.~\ref{fig:workflow}, and further describe each model below.

\begin{itemize}
\item The \textit{noise model} uses the {\tt BayesLine} algorithm, fully integrated into {\tt BayesWave}, to model the PSD $S_n(f)$ as a sum of a broadband spline and Lorentzians for spectral lines~\cite{Littenberg_2015}. The number and parameters of the spline control points and Lorentzians are marginalized over with an RJMCMC. In the workflow of Fig.~\ref{fig:workflow}, the noise RJMCMC is depicted with a blue box.
\item The \textit{CBC model} uses waveform templates derived within general relativity to model a CBC signal. The CBC sampler and its integration within {\tt BayesWave} are described in detail in~\cite{Wijngaarden:2022sah}. For this work, we use the {\tt IMRPhenomD}~\citep{Husa:2015iqa,Khan2016} waveform model with identical settings and priors as~\cite{Wijngaarden:2022sah}. The CBC likelihood is computed with the heterodyning method to reduce the computational cost~\cite{Cornish_2010,Cornish2021_HD}. In the workflow of Fig.~\ref{fig:workflow}, the CBC model parameters are split between the extrinsic MCMC that handles parameters $\theta_{\mathrm{ext}}$ (purple box) and the CBC MCMC that handles parameters $\theta^{\mathrm{cbc}}$ (pink box); see~\cite{Wijngaarden:2022sah} for details.
\item The \textit{glitch model} targets any incoherent non-Gaussian noise in the detectors by modeling such noise with a sum of sine-Gaussian wavelets. The number and morphology of the wavelets are not fixed \textit{a priori} but instead marginalized over with an RJMCMC. This model is typically used to target instrumental glitches, but can capture any excess power that is unaccounted for, including signals. In the workflow of Fig.~\ref{fig:workflow}, the glitch RJMCMC is depicted with a yellow box and its parameters are collectively denoted as $\theta_{\mathrm{w}}$.\footnote{For completeness we also mention the \textit{signal model} though it is not used in this study and thus not depicted in Fig.~\ref{fig:workflow}. The \textit{signal model} targets astrophysical signals by modeling coherent power between the detectors through a sum of sine-Gaussian wavelets. The number and morphology of the wavelets are again marginalized over with an RJMCMC. The \textit{signal model} is very similar to the \textit{glitch model}, only the wavelets are coherent rather than incoherent across the detector network. }
\end{itemize}

{\tt BayesWave} samples from the multidimensional posterior of all its models using a blocked Gibbs sampler. Each model is sampled iteratively using independent MCMC or RJMCMC samplers that update their corresponding block of parameters while keeping others parameters fixed. Each independent sampler proceeds for a given number of iterations, before returning the last set of parameters and switching to the next sampler. This procedure provides an efficient way of approximating the joint distribution of all parameters, provided that the parameters across blocks are not too correlated. The breakdown of parameters within blocks is described in~\cite{Wijngaarden:2022sah}, while~\cite{Hourihane:2022doe} shows that this procedure converges even when the parameters across blocks are correlated. 

{\tt BayesWave} can be used with any combination of these models, resulting in different analysis modes as shown in Table~\ref{tab:glossary}. The ``cleaning phase" is the standard {\tt BayesWave} mode used to compute median PSDs for analyses that use a point-estimate for the PSD. Data in each detector are analyzed separately with the \emph{noise model} and the \emph{glitch model}. The former targets the noise PSD, while the latter targets any excess non-Gaussian power (glitch or GW signal) to ensure that its presence does not affect PSD estimation. A posterior for the noise (and the excess power) is returned from which a median can be computed for further parameter estimation analyses such as~\cite{Abbott2019events,AbbottO3aEvents,LIGOScientific:2021djp}.
During the ``CBC+noise" analysis mode, data from all detectors are analyzed simultaneously. The GW signal is modeled with the \emph{CBC model} while the noise PSD in each detector is modeled with the \emph{noise model} and thus fully marginalized over. 
For comparison to the standard analysis of~\cite{Abbott2019events,AbbottO3aEvents,LIGOScientific:2021djp}, we also use the ``CBC+fixed" analysis mode where again the GW signal is modeled with the \emph{CBC model}, but now the noise PSD is fixed, typically acquired from a previous ``Cleaning Phase" analysis.

%----------------------------------------------
\subsection{Analysis description and data}
\label{sec:analysis}
%----------------------------------------------

\begin{table*}[]
\begin{tabular}{l@{\quad}|@{\quad} c @{\quad\quad} c @{\quad\quad} c @{\quad\quad} c @{\quad\quad} c @{\quad\quad}}
\multirow{2}{*}{Event} & \multirow{2}{*}{GPS Time (s)} & \multicolumn{1}{p{1.5cm}}{\centering{Segment\\length (s)}} & \multicolumn{1}{p{1.5cm}}{\centering{Sampling\\rate (Hz)}} & \multicolumn{1}{p{1.5cm}}{\centering{Detector\\network}}     & \multicolumn{1}{p{1.5cm}}{\centering{Network\\SNR}}   \\
\hline
GW150914           & 1126259462.420 & 4      & 2048  & H, L    & 24.4 \\
GW151012         & 1128678900.400 & 4      & 2048  & H, L & 10.0 \\
GW151226           & 1135136350.647 & 8      & 2048  & H, L    & 13.1 \\
GW170104           & 1167559936.599 & 8      & 2048  & H, L    & 13.0 \\
GW170729           & 1185389807.328 & 4      & 1024  & H, L, V & 10.2 \\
GW170809         & 1186302519.747 & 4      & 2048  & H, L, V & 12.4 \\
GW170814           & 1186741861.527 & 4      & 2048  & H, L, V & 15.9 \\
GW170818           & 1187058327.082 & 4      & 2048  & H, L, V & 11.3 \\
GW170823           & 1187529256.518 & 4      & 2048  & H, L    & 11.5 \\
GW190408\_181802   & 1238782700.286 & 8      & 2048  & H, L, V & 14.7 \\
GW190412           & 1239082262.170 & 8      & 4096  & H, L, V & 18.9 \\
GW190421\_213856   & 1239917954.260 & 4      & 1024  & H, L    & 10.6 \\
GW190503\_185404   & 1240944862.298 & 4      & 2048  & H, L, V & 12.1 \\
GW190512\_180714  & 1241719652.419 & 8      & 2048  & H, L, V & 12.3 \\
GW190513\_205428   & 1241816086.747 & 4      & 2048  & H, L, V & 12.3 \\
GW190517\_055101   & 1242107479.838 & 4      & 2048  & H, L, V & 10.2 \\
GW190519\_153544   & 1242315362.398 & 4      & 1024  & H, L, V & 12.1 \\
GW190521\_030229  & 1242442967.460 & 4      & 512   & H, L, V & 14.3 \\
GW190521\_074359   & 1242459857.466 & 4      & 1024  & H, L    & 24.4 \\
GW190602\_175927   & 1243533585.089 & 4      & 1024  & H, L, V & 12.1 \\
GW190620\_030421   & 1245035079.311 & 4      & 1024  & L, V    & 10.9 \\
GW190630\_185205   & 1245955943.180 & 4      & 2048  & L, V    & 15.6 \\
GW190701\_203306   & 1246048404.580 & 4      & 1024  & H, L, V & 11.6 \\
GW190706\_222641   & 1246487219.346 & 4      & 2048  & H, L    & 12.4 \\
GW190707\_093326   & 1246527224.169 & 16     & 4096  & H, L    & 13.0 \\
GW190708\_232457   & 1246663515.384 & 8      & 4096  & L, V    & 13.1 \\
GW190720\_000836   & 1247616534.707 & 16     & 4096  & H, L, V & 11.7 \\
GW190728\_064510  & 1248331528.534 & 16     & 8192  & H, L, V & 13.6 \\
GW190828\_063405   & 1251009263.800 & 4      & 2048  & H, L, V & 16.0 \\
GW190828\_065509   & 1251010527.890 & 8      & 2048  & H, L, V & 11.1 \\
GW190910\_112807   & 1252150105.324 & 4      & 1024  & L, V    & 13.4 \\
GW190915\_235702   & 1252627040.692 & 8      & 2048  & H, L, V & 13.1 \\
GW191109\_010717   & 1257296855.217 & 4      & 1024  & H, L    & 17.3 \\
GW191129\_134029   & 1259070047.197 & 16     & 8192  & H, L    & 13.1 \\
GW191204\_171526   & 1259514944.092 & 8      & 4096  & H, L    & 17.5 \\
GW191215\_223052   & 1260484270.334 & 8      & 2048  & H, L, V & 11.2 \\
GW191216\_213338 & 1260567236.472 & 16     & 8192  & H, V    & 18.6 \\
GW191222\_033537   & 1261020955.124 & 8      & 1024  & H, L    & 12.5 \\
GW191230\_180458   & 1261764316.407 & 4      & 2048  & H, L, V & 10.4 \\
GW200112\_155838   & 1262879936.091 & 4      & 2048  & L, V    & 19.8 \\
GW200128\_022011   & 1264213229.901 & 4      & 1024  & H, L    & 10.6 \\
GW200129\_065458   & 1264316116.433 & 8      & 2048  & H, L, V & 26.8 \\
GW200202\_154313   & 1264693411.556 & 16     & 8192  & H, L, V & 10.8 \\
GW200208\_130117   & 1265202095.950 & 4      & 1024  & H, L, V & 10.8 \\
GW200219\_094415   & 1266140673.197 & 4      & 2048  & H, L, V & 10.7 \\
GW200224\_222234   & 1266618172.402 & 4      & 1024  & H, L, V & 20.0 \\
GW200225\_060421  & 1266645879.396 & 8      & 2048  & H, L    & 12.5 \\
GW200302\_015811   & 1267149509.516 & 8      & 2048  & H, V    & 10.8 \\
GW200311\_115853   & 1267963151.300 & 4      & 2048  & H, L, V & 17.8 \\
GW200316\_215756 & 1268431094.158 & 16     & 4096  & H, L, V & 10.3 \\
\hline
\hline
\end{tabular}
\caption{Table of events and analysis settings. The columns in order provide the event name, its geocenter GPS time, the analysis segment length, the analysis sampling rate, the detector network (H: LIGO Hanford, L: LIGO Livingston, V: Virgo), and network SNR as reported in the GW Open Science Center. For all events, we use a lower frequency cutoff of $f_{\rm low} = 20$ Hz.}
\label{tab:settings}
\end{table*}

We analyze the BBH events from GWTC-3 with SNR above 10, excluding events that contain neutron stars to limit computational cost. A list of events and analysis settings are provided in Table~\ref{tab:settings}. In order to examine the effect of uncertainty on the noise PSD on inference, we perform two analyses on each event, which we label \emph{sequential} and \emph{marginalized}.
\begin{enumerate}
    \item The \emph{sequential} method is constructed to closely follow the traditional fixed PSD analysis. First, we use the ``cleaning phase" mode (see Table~\ref{tab:glossary}) to obtain a posterior for the PSD. From that posterior, we compute the per-frequency-bin median PSD. Second, we use the ``CBC+fixed" mode with the previously computed median PSD as input.
    \item The \emph{marginalized} method includes a single analysis of the data with the ``CBC+noise" mode (see Table~\ref{tab:glossary}) where the CBC and PSD are inferred simultaneously. 
\end{enumerate}

%%%%%%%%%%%%%%%%%%%%%%%%%%%%%%%%%%%%%%%%%%
\section{Results on GWTC-3 Events}
\label{sec:analyses}
%%%%%%%%%%%%%%%%%%%%%%%%%%%%%%%%%%%%%%%%%%

In this section we present the results of the \emph{sequential} and \emph{marginalized} noise methods for each event. We begin with an in-depth discussion of GW150914~\cite{LIGOScientific:2016aoc} and then present results for all events.

%----------------------------------------
\subsection{GW150914}
\label{sec:gw150914}

\begin{figure*}[]
    \centering
    \includegraphics[width=0.98\textwidth]{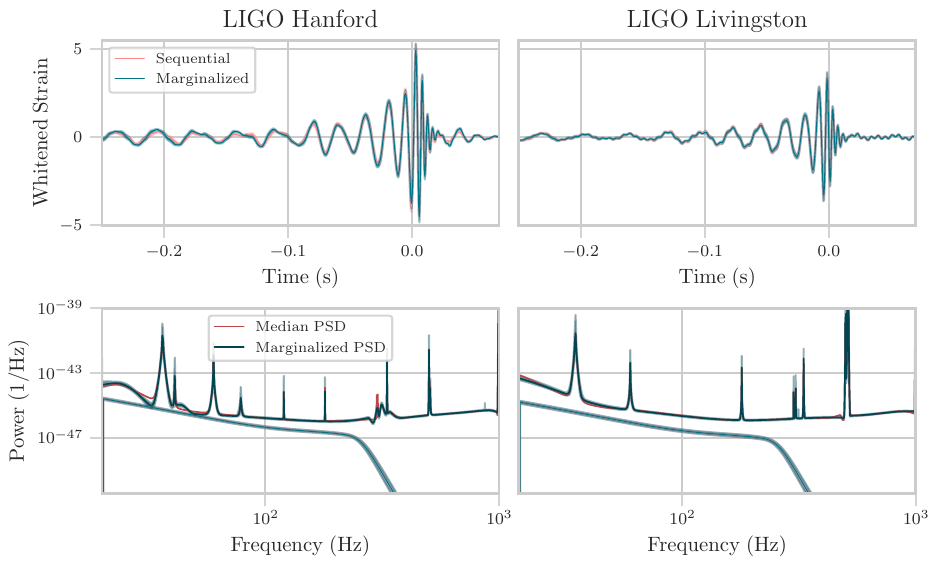}
    \caption{ Whitened time-domain reconstruction (top) and spectrum (bottom) of GW150914 in LIGO Hanford (left) and LIGO Livingston (right). The median and 90\% credible interval for the signal are shown in pink and teal for the sequential and the marginalized analyses, respectively. The red line gives the median PSD used for the sequential analysis, while the median and 90\% credible interval for the PSD from the marginalized analysis are in dark blue. Time is with respect to GPS 1126259462.42.}
    \label{fig:150914_waveforms}
\end{figure*}

We show the time- and frequency-domain reconstructions of GW150914 in Fig.~\ref{fig:150914_waveforms} with both methods. We find that the recovered waveforms agree in median and uncertainty; despite the marginalized method adding free parameters to the analysis, the resulting reconstructions have comparable widths. The marginalized and median PSDs agree quite well overall, with minor deviation around the low-frequency spectral lines in LIGO Hanford.

\begin{figure}[]
    \centering
    \includegraphics[width=0.48\textwidth]{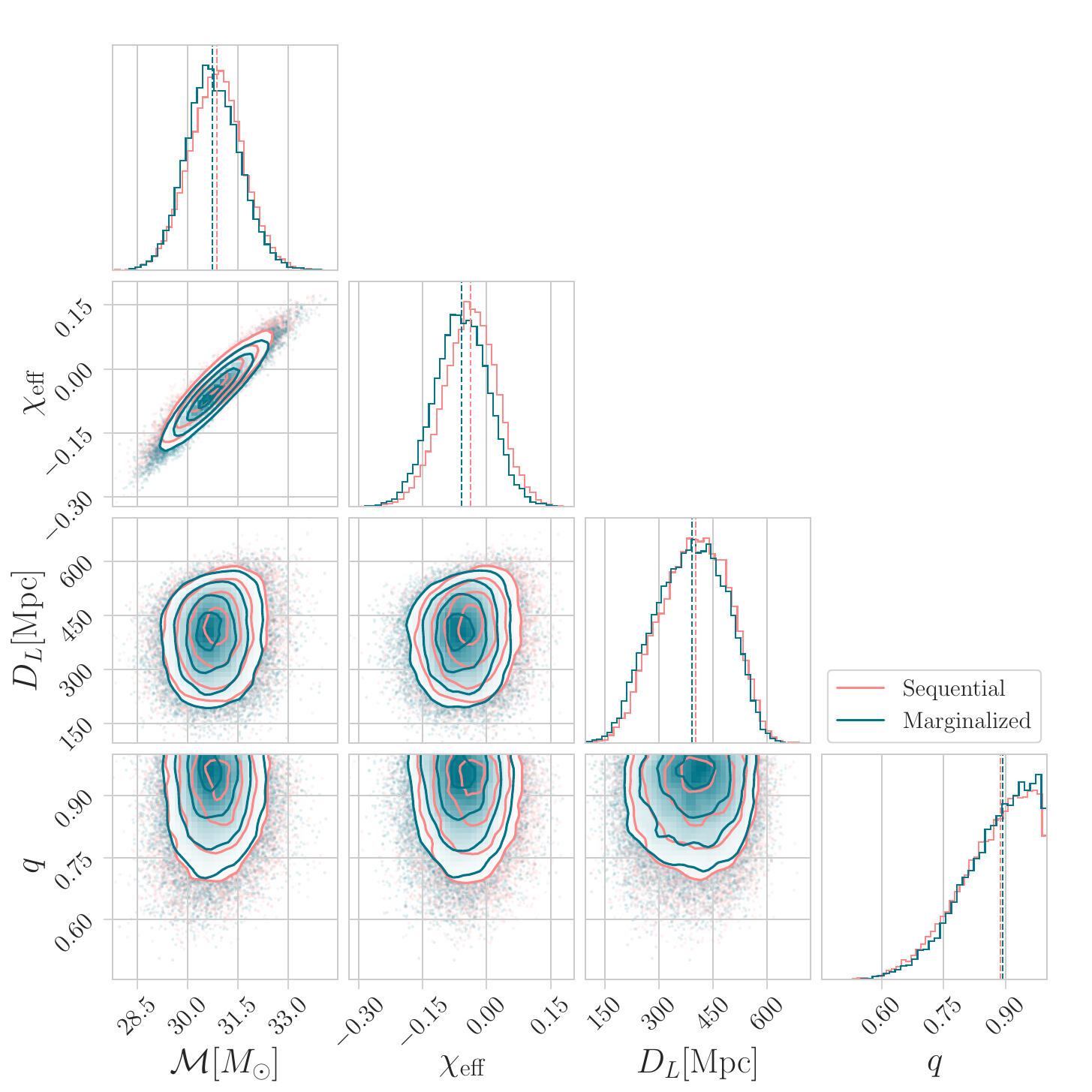}
    \caption{One- and two-dimensional marginalized posterior for the detector-frame chirp mass $\Mc$, the effective spin $\chieff$, the luminosity distance $\DL$, and the mass ratio $q$ of GW150914. We show results obtained with the sequential and the marginalized noise estimation methods in pink and teal, respectively. The methods give consistent results to within the statistical uncertainty.}
    \label{fig:150914posteriors}
\end{figure}

We then consider the posteriors for select CBC parameters. In Fig.~\ref{fig:150914posteriors} we show one- and two-dimensional marginalized posteriors for the detector-frame chirp mass,
\begin{equation}
    \Mc \equiv \frac{(m_1m_2)^{3/5}}{(m_1+m_2)^{1/5}};
\end{equation}
the effective spin,
\begin{equation}
    \chieff \equiv \frac{m_1\chi_1\cos\theta_1+m_2\chi_2\cos\theta_2}{m_1+m_2},
\end{equation}
which is a relatively well-measured spin parameter~\cite{Ng:2018neg}; the mass ratio $q\equiv m_2/m_1<1$; and the luminosity distance $\DL$. In the above equations, $m_i$, $\chi_i$, and $\cos\theta_i$ with $i\in\{1,2\}$ are the component masses, spin magnitudes, and angles between the spin and the Newtonian orbital angular momentum, respectively. As with the waveforms, the parameter posteriors are consistent within the statistical measurement uncertainty. For all parameters, the median appears shifted to slightly larger values but the effect is small, with only a $0.4\%$ shift in the chirp mass. The marginalized method additionally recovers $\sim 1\%$ higher network SNR, which is a minimal difference possibly attributable to the slightly closer distance posterior.

%----------------------------------------
\subsection{Remaining events}
\label{sec:otherevents}

\begin{figure*}
    \centering
    \includegraphics[width=0.92\textwidth]{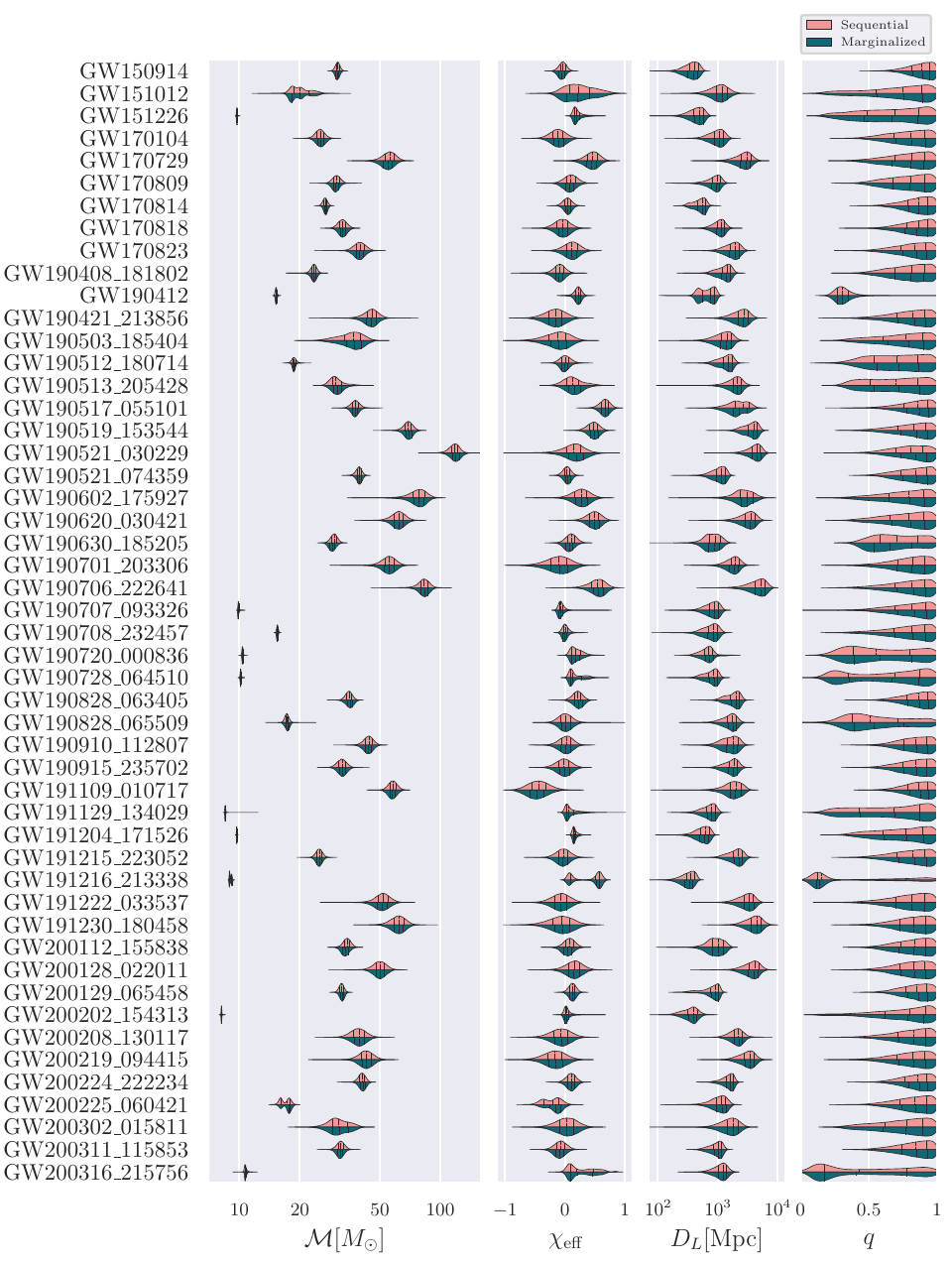}
    \caption{Violin plots of the posterior probability density of the sequential (pink) compared to the marginalized (teal) noise estimation method for the detector-frame chirp mass ${\cal{M}}$, effective spin $\chi_{\mathrm{eff}}$, luminosity distance $D_{L}$, and mass ratio $q$ of each event. The dashed and dotted lines depict the quartiles for each posterior.}
    \label{fig:violinPlot}
\end{figure*}

We present the one-dimensional marginalized posterior for $\Mc$, $\chieff$, $\DL$, and $q$ for the entire set of GWTC-3 events in Fig.~\ref{fig:violinPlot}. While we find small variation in posterior position and width between the marginalized and the sequential methods, overall there is broad agreement. 

\begin{figure}[!ht]
    \centering
    \includegraphics[width=0.48\textwidth]{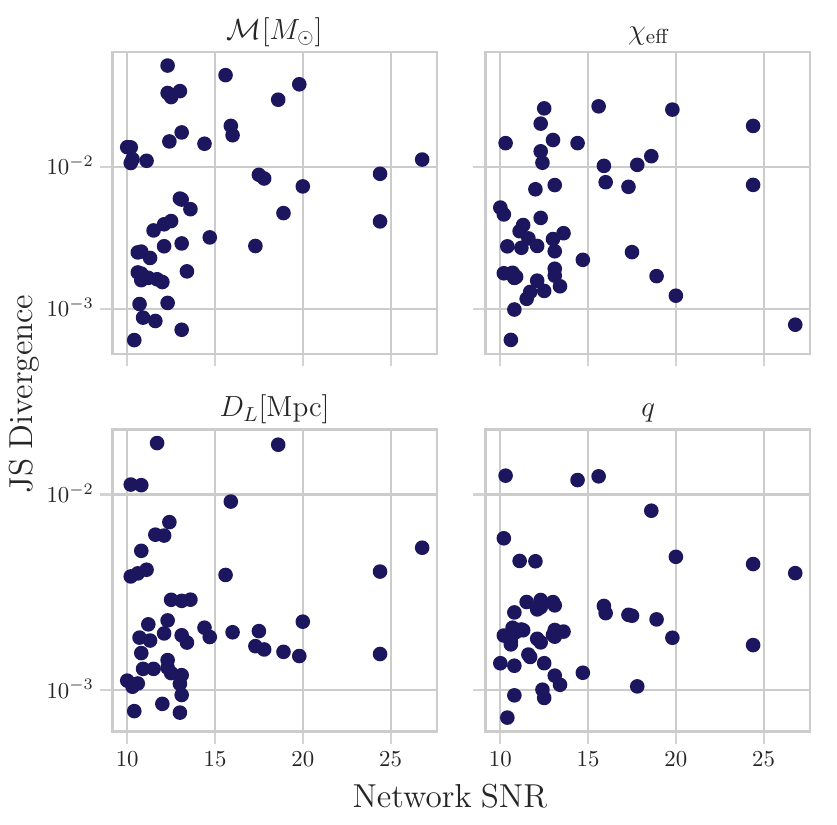}
    \caption{Jensen-Shannon divergence (JSD) between the sequential and marginalized parameter posteriors as a function of the network matched-filter SNR for $\Mc$, $\chi_{\rm eff}$, $D_L$, and $q$. Each dot corresponds to one event from Table~\ref{tab:settings}. Across all events we find JSDs $\leq 0.05$.}
    \label{fig:js}
\end{figure}

To quantify the difference between the methods, we compute the Jensen-Shannon divergence (JSD)~\cite{Lin1991} between the sequential and marginalized posteriors. This summary statistic describes the similarity of two distributions and ranges from 0 for identical distributions to~1 for dissimilar distributions. In Fig.~\ref{fig:js}, we show the JSD for the marginalized $\Mc$, $\chi_{\rm eff}$, $D_L$, and $q$ posteriors as a function of the network matched-filter SNR as reported in GWOSC~\cite{Vallisneri:2014vxa,LIGOScientific:2019lzm}. The average JSD is $0.006$ across all events and parameters and the JSD is smaller than $0.05$ for all except one, indicating overall agreement between the methods~\cite{Abbott2019events}. The largest JSD is 0.052, for the chirp mass of GW190512\_180714. The magnitude arises from a sizeable posterior width discrepancy, which is further discussed below and is shown in Appendix~\ref{sec:outliers} to depend on the analysis low frequency cutoff. Overall, the difference between the marginalized and sequential posteriors is below the difference in posteriors induced by choice of waveform model~\cite{Abbott2019events}.  In addition, the JSD shows minimal trend with SNR with the exception of a slight trend of increasing chirp mass JSD with SNR. However, the underdensity of events at high SNR makes it premature to confirm a relation.

\begin{figure}[!ht]
    \centering
    \includegraphics[width=0.48\textwidth]{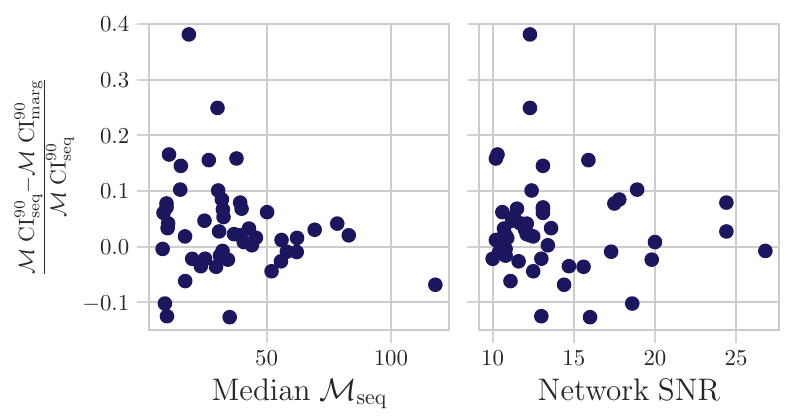}
    \caption{Fractional difference between the width of the 90\% credible interval of the sequential and marginalized methods as a function of the network matched-filter SNR and the median posterior $\Mc$. We find no overall trend.}
    \label{fig:snr_vs_width}
\end{figure}

Further, we examine the chirp mass posterior width in Fig.~\ref{fig:snr_vs_width}, where we plot the relative difference between the two methods as a function of the posterior median (i.e., the event chirp mass) and the SNR. We find no significant trends between these parameters, concluding that the effect of marginalizing over PSD uncertainty does not  depend on the chirp mass or signal strength. Indeed, a linear fit of the data of Fig.~\ref{fig:snr_vs_width} returns a slope consistent with zero. The biggest outlier is GW190512\_180714, for which we obtain a $\sim40\%$ \textit{decrease} in posterior width using the marginalized PSD. Upon further testing, reducing the lower frequency limit to $f_{\rm low} = 16$\,Hz from $f_{\rm low} = 20$\,Hz reduces these differences. We apply the same test to another event with a large posterior width discrepancy, GW190513\_205428 ($\sim 25\%$ decrease) and reach the same conclusion. We continue to use the standard settings for the remaining analyses, but present the $f_{\rm low} = 16$\,Hz results in Appendix~\ref{sec:outliers}.

%%%%%%%%%%%%%%%%%%%%%%%%%%%%%%%%%%%%%%%%%%%%%%%%%%
\section{Toy model of PSD variation}
\label{sec:toymodel}
%%%%%%%%%%%%%%%%%%%%%%%%%%%%%%%%%%%%%%%%%%%%%%%%%%

%
\begin{figure}[]
    \centering
    \includegraphics[width=0.48\textwidth]{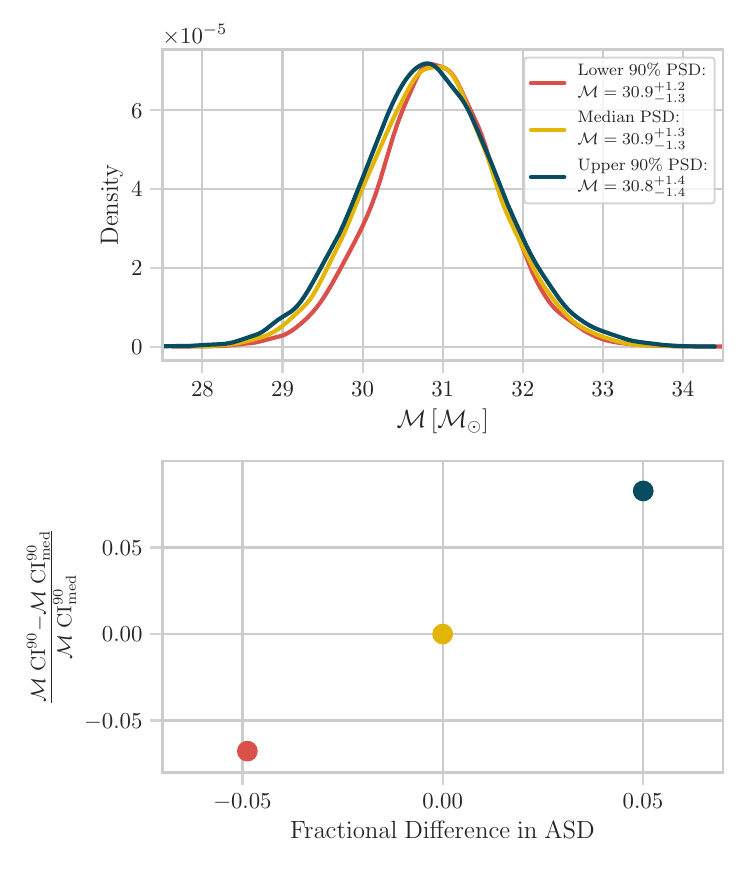}
    \caption{Top: marginalized detector-frame chirp mass posterior for GW150914 with the ``CBC+fixed" analysis using the median, lower 90\%, and upper 90\% estimate for the PSDs. Bottom: fractional difference between the 90\% credible interval widths of the chirp mass posteriors from the top panel against the fractional difference between amplitude spectral density and the median at 100 Hz.}
    \label{fig:150914_shifted_psds}
\end{figure}

The results of Sec.~\ref{sec:analysis} suggest that the sequential noise estimation method results in CBC parameter posteriors that are qualitatively similar to the marginalized method. The main difference between the two is that the former uses the median of the PSD posterior, while the latter uses the full posterior. To further explore the difference between the two, we consider what happens if we instead had selected the upper or lower 90\% estimate of the PSD with the sequential method. A larger noise PSD should lead to underestimation of the signal SNR, and as a consequence, overestimation of the posterior uncertainty. Similarly, a lower PSD should lead in more narrow posteriors. In effect, we expect a correlation between the noise posterior height and the CBC parameter posterior width.

To explore this behavior in the data, we revisit the ``CBC+fixed" analysis for GW159014 using not only the median PSD, but also the lower and upper 90\% credible interval PSDs. The chirp mass posterior for each analysis is plotted in the top panel of Fig.~\ref{fig:150914_shifted_psds}; they show the expected posterior broadening and narrowing but also a small shift in the posterior median. The bottom panel of Fig.~\ref{fig:150914_shifted_psds} shows the width of the 90\% credible interval for $\mathcal M$ as a function of the height of the amplitude spectral density (i.e., the square root of the PSD that is proportional to the SNR and thus the posterior width), given as fractional difference to the median at 100 Hz. Though with three points it is premature to conclude that the trend is linear, it is still evident that the uncertainty increases as the PSD increases (and the SNR decreases). In the case of GW159014, we also find an increase in the posterior median with decreased PSD height. However, tests on other events confirm the trend of the bottom panel of Fig.~\ref{fig:150914_shifted_psds} regarding the posterior width, but no trend regarding the median. This suggests that the median shift is not a generic feature and further confirms the expected correlation between the posteriors.

\begin{figure}[]
    \centering
    \includegraphics[width=0.48\textwidth]{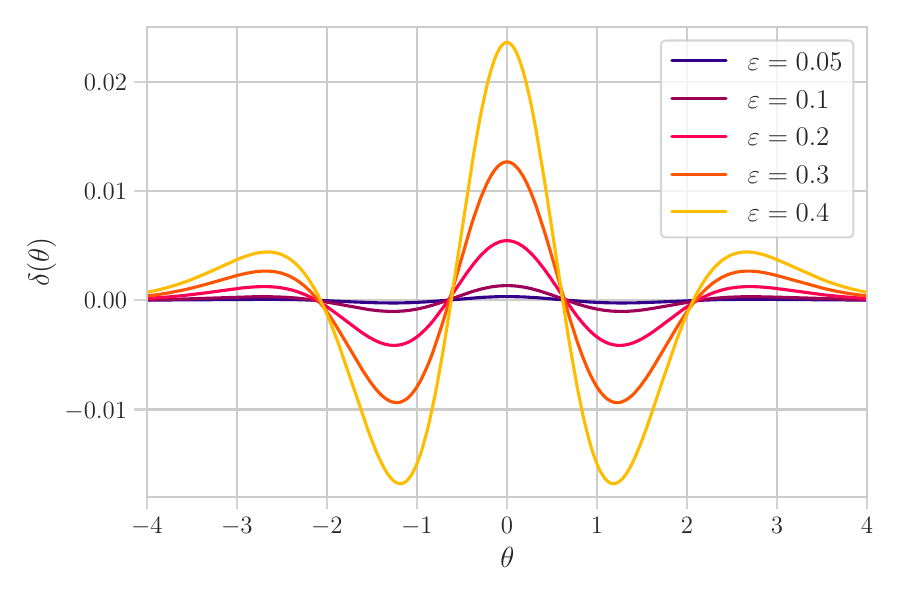}
    \caption{Plot of $\delta(\theta)$, Eq.~\eqref{eq:delta}, the difference between $\mathcal N(\theta|0,1)$ and $\int_{-\varepsilon}^\varepsilon \mathcal N(\theta|0,1+x)dx$ for various values of $\varepsilon$. A Taylor expansion shows that $\delta(\theta)\sim \varepsilon^2$ at $\theta=0$.}
    \label{fig:dtheta}
\end{figure}

We further explore a toy model to illuminate this behavior. When marginalizing over PSD uncertainty, each step in the CBC MCMC uses a different PSD that is drawn from the PSD posterior. Overall, the sampler therefore uses a range of PSDs that can be either higher or lower than the median, and thus the CBC posterior width of each sample will be either increased or decreased compared to the median. As a result, the CBC posterior that results from marginalizing over PSD parameters will be a collection of samples from posteriors with different standard deviations. Simplifying to the case of a Gaussian posterior, this effect is comparable to a normal distribution integrated over a range of standard deviations. The difference between the posteriors from using a fixed median PSD (the sequential method) versus the full PSD posterior (the marginalized method) can then be approximated by the difference between a normal distribution and one integrated over a range of standard deviations,
\begin{equation}
    \delta(\theta) \equiv \frac{1}{2\varepsilon}\int_{-\varepsilon}^\varepsilon \mathcal N\big(\theta|\mu,(1+x)\sigma\big)dx - \mathcal N(\theta|\mu,\sigma)\,, \label{eq:delta}
\end{equation}
where $\theta$ is the parameter of interest, $\mu$ is the mean of the posterior, $\sigma$ is the standard deviation when using the median PSD, and $\varepsilon$ encodes the relative PSD uncertainty. Here we choose $\varepsilon$ symmetric about $\sigma$, which arises from the expectation (and observation, as shown in Fig.~\ref{fig:150914_shifted_psds}) that if the ASD uncertainty is Gaussian and thus symmetric about the median, then the resulting range of posterior widths will also be roughly symmetric. Figure~\ref{fig:dtheta} shows $\delta(\theta)$ for $\mu=0$, $\sigma=1$, and several values of $\varepsilon$. A Taylor expansion around $\varepsilon=0$ shows that the difference between the two distributions peaks at their mean and goes as $\delta(\theta)\sim \varepsilon^2$. This scaling suggests that a small PSD uncertainty (and thus a small $\varepsilon$) results in a suppressed effect in the parameter posteriors as it is a higher order effect. Indeed as shown in Fig.~\ref{fig:150914_shifted_psds}, the typical PSD uncertainty at current sensitivities is $\varepsilon \sim 0.05$, which is much smaller than the typical width of the posteriors.
%

%%%%%%%%%%%%%%%%%%%%%%%%%%%%%%%%%%%%%%%%%%
\section{Conclusion}\label{sec:conclusion}
%%%%%%%%%%%%%%%%%%%%%%%%%%%%%%%%%%%%%%%%%%

In this study, we assessed the impact of assuming a known and fixed PSD during parameter estimation for CBCs. We compared two methods, a \emph{sequential} method that first computes the PSD posterior and then uses the median to estimate the posterior for the CBC parameters, and a \emph{marginalized} method that simultaneously infers the posterior for the PSD and the CBC parameters. We recovered similar parameter posteriors, in agreement with previous results~\cite{Talbot2020,Biscoveanu2020}. In particular, the difference in the chirp mass posterior median (width) between the two methods is $0^{+1.4}_{-1.6}\%$ ($2^{+16}_{-13}\%$) across all events. We found no trends between these differences and the event chirp masses and network SNRs.
The marginalized method involves simultaneous modeling of more parameters and thus comes with a greater computational cost than the sequential method. This increase is smaller for high-mass events, but can reach an increase of up to $\sim1.5$ in computing time for low-mass events.

The simultaneous modeling of the CBC signal and the noise PSD allows us to marginalize over uncertainty in the latter. Compared to the method explore in~\cite{Biscoveanu2020}, our method involves a single parameter estimation analysis and can account for potential correlations between the CBC signal and the PSD model parameters. Additionally, our method enables estimation of the Bayesian evidence for the marginalized PSD analysis and thus computation of the Bayes factor between different models. However, correctly interpreting the Bayes factor within the context of the chosen priors of the PSD model is a nontrivial task and lies outside the scope of this paper.

We explained our results within the context of a toy model that considers the impact of a systematic increase or decrease of the PSD. In the case of simple Gaussian distributions, the difference between the posterior from the median PSD and the one from the full PSD posterior is of higher order in the PSD uncertainty. Thus the \emph{sequential} method that is based on the \emph{median} PSD already results in posteriors that are very similar to the \emph{marginalized} method.
We conclude that at current detector sensitivities, PSD uncertainty is subdominant compared to other sources of uncertainty, especially when the median on-source PSD is used.

%%%%%%%%%%%%%%%%%%%%%%%%%%%%%%%%%%%%%%%%%%
\section{Acknowledgments}
%%%%%%%%%%%%%%%%%%%%%%%%%%%%%%%%%%%%%%%%%%

This work is supported by National Science Foundation grant No. PHY-1852081 as part of the LIGO Caltech REU Program.
This research has made use of data, software and/or web tools obtained from the Gravitational Wave Open Science Center~\cite{gwosc}, a service of LIGO Laboratory, the LIGO Scientific Collaboration and the Virgo Collaboration.
Virgo is funded by the French Centre National de Recherche Scientifique (CNRS), the Italian Istituto Nazionale della Fisica Nucleare (INFN) and the Dutch Nikhef, with contributions by Polish and Hungarian institutes.
This material is based upon work supported by NSF's LIGO Laboratory which is a major facility fully funded by the National Science Foundation.
The authors are grateful for computational resources provided by the LIGO Laboratory and supported by NSF Grants No. PHY-0757058 and No. PHY-0823459. 
S.H. and K.C. were supported by NSF Grant No. PHY-2110111.
Software: {\tt GWPY}~\cite{duncan_macleod_2020_3598469}, {\tt MATPLOTLIB}~\cite{Hunter:2007}, {\tt CORNER}~\cite{corner}.

\appendix

%%%%%%%%%%%%%%%%%%%%%%%%%%%%%%%%%%%%%%%%%%
\section{Outlier follow-up}\label{sec:outliers}
%%%%%%%%%%%%%%%%%%%%%%%%%%%%%%%%%%%%%%%%%%

\begin{figure}[]
    \centering
    \includegraphics[width=0.48\textwidth]{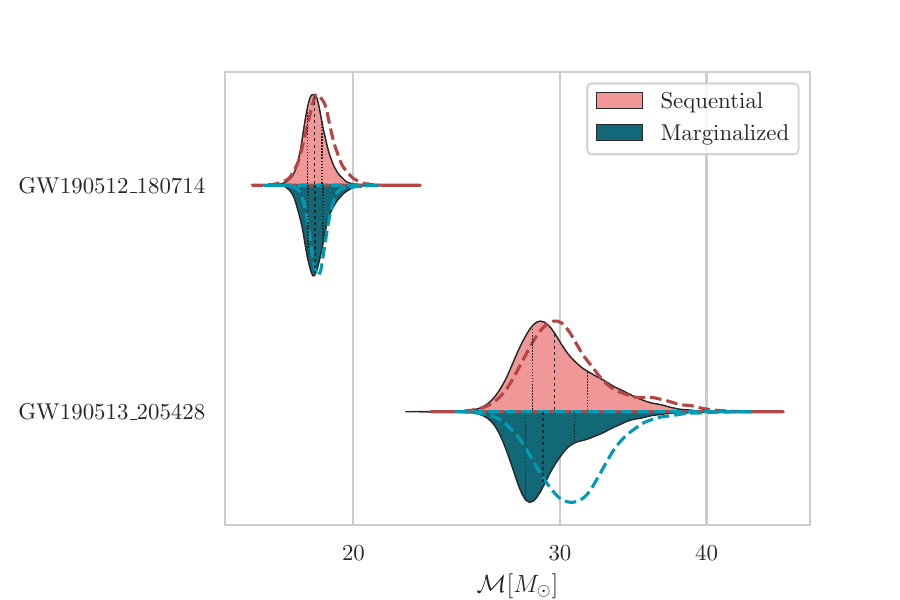}
    \caption{Detector-frame chirp mass posterior comparison between the sequential (pink) and marginalized (teal) analyses for GW190512\_180714 and GW190513\_205428 using $f_{\rm low} = 16$\,Hz. These events have larger discrepancies between the posteriors when using $f_{\rm low} = 20$\,Hz, presented in dashed for comparison.}
    \label{fig:flow16_events}
\end{figure}

We revisit the two events with the largest differences in parameter posteriors between the marginalized and sequential method, GW190512\_180714 and GW190513\_205428.
In the analyses with the standard settings presented in Sec.~\ref{sec:otherevents}, we found fractional differences in the chirp mass 90\% posterior width of 38.1\% and 24.9\%, respectively. Among the various parameter settings, reducing the lower frequency cutoff from $f_{\rm low} = 20$\,Hz to $f_{\rm low} = 16$\,Hz significantly mitigates the observed differences. We compare the different $\Mc$ posteriors for GW190512\_180714 and GW190513\_205428 in Fig.~\ref{fig:flow16_events}.  Using $f_{\rm low} = 16$\,Hz reduces the posterior width differences to -7.1\% and 11.3\%, respectively. Similarly, the JSDs are reduced from 0.052 and 0.033 to 0.0025 and 0.015 for each event, respectively. This follow-up analysis confirms that the differences between the posteriors of these events goes down with smaller lower frequency cutoffs, and suggests that analysis settings can sometimes significantly influence parameter estimation results.

\bibliography{references.bib}

\end{document}